Comments on the Quantum Afterburner

Elias P. Gyftopoulos
Massachusetts Institute of Technology
77 Massachusetts Avenue
Cambridge, Massachusetts 02139

Michael R. von Spakovsky
Virginia Polytechnic Institute and State University
Energy Management Institute
Blacksburg, Virginia 24061-0238

A process has been proposed to increase the efficiency of an ideal Otto cycle via a quantum heat engine that has no cooler reservoir. We show that such a process is not feasible.

PACS numbers: 05.70.Ln, 03.65.-w, 05.30.-d

In two recent publications, Scully proposes to extract work from a single bath via quantum negentropy [1], thus improving the efficiency of an ideal Otto cycle engine [2] by exploiting presumed properties of spins. In our view, the proposed process does not and cannot result in any efficiency improvement because it violates the *complete* laws of physics. A brief summary of the complete laws of physics is given by Gyftopoulos [3, 4]. They are based on two interrelated developments that occurred over the past three decades. One is an exposition of thermodynamics as a nonstatistical science that applies to all systems (both macroscopic and microscopic, including a one spin system), and to all states (both thermodynamic equilibrium and not thermodynamic equilibrium). In the new exposition, entropy is proven to be a nonstatistical property of each constituent of a system, in the same sense that inertial mass is a property of each constituent of a system. The second development is a unified quantum theory of mechanics and thermodynamics. This theory extends quantum principles to thermodynamic phenomena, and thermodynamic principles to quantum phenomena without use of statistical probabilities.

In agreement with Scully, and without loss of generality, we assume temperature and pressure conditions along the Otto cycle such that the working fluid behaves as a perfect gas. Perfect gas behavior implies that each molecule of the working fluid is a separable and uncorrelated subsystem, and has specific heats independent of temperature. Furthermore, because the translational degrees of freedom of a molecule are not coupled to its spin, each molecule can be regarded as a composite consisting of two separable and uncorrelated subsystems, one defined by the molecular translational degrees of freedom, and the other by the spin. Finally, and also in agreement with Scully, we assume that in state 4 of the Otto cycle (Fig. 2, Ref. [2]) both the molecule and the spin are in states in mutual stable equilibrium with the environmental reservoir, that is, the available energy (exergy) of both the molecule and the spin subsystems are equal to zero. Thus, the following well-founded conclusions can be asserted:

(i) An ideal (reversible) Otto cycle, such as the one discussed by Scully, has a thermodynamic efficiency (not thermal efficiency) equal to 100%; this efficiency includes the contributions of both subsystems, that is, the one defined by the translational degrees of freedom, and the other by the spin. So the thermodynamic efficiency cannot be improved.

(ii) In the absence of either lasing and/or masing, each of the two subsystems of a molecule are in mutual stable equilibrium at each state of the Otto cycle. As such, the composite of these two subsystems is characterized by a density operator $\rho^0 = \rho_t^0 \times \rho_s^0$, where $\rho_t^0$ and $\rho_s^0$ are the density operators associated with the translational degrees of freedom and the spin, respectively, at each pair of temperature and pressure of the state of the Otto cycle. In matrix form, both $\rho_t^0$ and $\rho_s^0$ are diagonal in the energy representation of the corresponding subsystem. In addition, and more importantly, it is noteworthy that according to the complete laws of physics, each of the density operators $\rho_t^0$ and $\rho_s^0$ is representable by one and only one homogeneous ensemble of identical systems, identically prepared [3, 4]. An ensemble representing a density operator $\rho$ is homogeneous if and only if in no conceivable way it can be rearranged into subensembles each of which is characterized by a density operator $\rho_i$ such that

$$\rho = \Sigma_i \, \alpha_i \rho_i \qquad \text{and} \qquad \Sigma_i \, \alpha_i = 1$$

where $\rho_i \neq \rho$ and $\alpha_i$ is the fraction of members of the ensemble that corresponds to $\rho_i$. Said differently, an ensemble is homogeneous if and only if $\rho$ operationally (as opposed to numerically) cannot be construed as a statistical average of different $\rho_i$'s because if it can then it is not subject to the complete laws of physics.

(iii) In particular, the density operator that corresponds to the spin subsystem at temperature $T_4$ is not a statistical mixture of hot (spin up), and cold (spin down) parts as shown in Fig. 1b of Ref. [1] because $\rho_s^0$ must be represented by a homogeneous ensemble and, therefore, physically cannot be regarded as a statistical average of two projectors, one for spins up and the other for spins down. In addition, and perhaps more importantly, the purely spin up and spin down projectors have ordinary temperatures $T_{su} = 0$ and $T_{sd} = 0$, respectively, but thermodynamic measures of temperature $1/T_{su} = -\infty$ and $1/T_{sd} = +\infty$, and the laws of thermodynamics require that energy and entropy flow from a part at $1/T$ negative to a part $1/T$ positive [5]. So the spin configurations depicted in Fig. 1b of Ref. [2] violate the complete laws of physics.

(iv) Selectively, the zero available energy of the spin at temperature $T_4$ can be raised to a nonzero available energy state by lasing and/or masing, and thus get work out in addition to the work done in the course of the ideal Otto cycle as the spin returns to a state in mutual stable equilibrium with the subsystem defined by the translational degrees of freedom. At best, however, the additional work will be equal to the available energy expended in lasing and/or masing, and usually less than the latter because of entropy generated spontaneously by irreversibility in the course of increasing and decreasing the available energy of the spin.

In view of the preceding analyses, we regret to conclude that the quantum afterburner does not and cannot increase the thermodynamic efficiency of an engine undergoing an ideal Otto cycle.